\documentclass[%
 preprint, linenumbers,
nofootinbib,
 amsmath,amssymb,
 aps, physrev,
]{revtex4-2}

\let\oldfootnote\footnote
\renewcommand{\footnote}[1]{%
    \begingroup%
    \linespread{1}%    % <- linespread for footnote: 1, 1.1, 1.2 etc
    \oldfootnote{#1}%
    \endgroup%
}

\usepackage{graphicx}% Include figure files
\usepackage{dcolumn}% Align table columns on decimal point
\usepackage{bm}% bold math
\usepackage{hyperref}% add hypertext capabilities
\usepackage{xcolor}

\begin{document}
\nolinenumbers
%\preprint{APS/123-QED}

\title{\textbf{Atomistic Simulations Reveal the Need to Reassess Standard Thermodynamic Models of Coherent Precipitates} 
}% 

\author{Anas Abu-Odeh}
 \email{Contact author: anas.abu-odeh@nist.gov}
\author{James Warren}%
\affiliation{%
Materials Measurement Laboratory, National Institute of Standards and Technology, Gaithersburg, MD, USA}%

\date{\today}% It is always \today, today,

\begin{abstract}
Accurate models of precipitation kinetics are essential to control and design structural materials. These models are highly sensitive to the thermodynamic description of precipitates. We use atomistic simulations of a model Fe-Cr system to assess two commonly used assumptions in the thermodynamic modeling of coherent precipitates: that elastic effects can be neglected for systems with a small lattice misfit and that size effects can be neglected for low levels of supersaturation. We find that these assumptions cannot be maintained for an accurate description of interfacial equilibrium, even when lattice misfits are below 1 \% and supersaturation values are below 1 \%. Additionally, we find a surprising trend at large precipitate radii that suggests the importance of higher-order effects that are commonly neglected. The results and insights from this study highlight the need to revisit current approaches in modeling solid-state precipitation.
\end{abstract}

\maketitle
%cccccbhdjucvcvjvtvcdnetdjghbnhfbdrebkntfekvi
\newpage

%\tableofcontents

\section{Introduction}

Solid-state precipitation influences the mechanical properties of structural materials. This can either be beneficial, for example by increasing the strength of materials \cite{Ardell1985}, or detrimental, for example by reducing the fracture toughness of materials \cite{Hamano1993}. The control of processing conditions to promote or suppress precipitates is thus crucial to the design of structural materials with desired mechanical properties. There is thus a  need for accurate models of phase transformation kinetics.

Phase transformation kinetics depend sensitively on   the thermodynamic description of precipitates. This is exemplified by a recent study by Tavenner et al. \cite{Tavenner2024} where it was shown that incorporating an experimental uncertainty of approximately 20 \% in the interfacial free energy for the modeling of coherent precipitation kinetics in a Ni-Al alloy resulted in a large range of uncertainty of phase fraction and mean radius evolution over time. This result is particularly interesting when considering that the Ni-Al system is one where the precipitation kinetics and thermodynamics has been studied for many decades (see for example Refs. \cite{Ardell1995,Ardell2021} and references therein). Most alloy systems do not have the privilege of an exhaustive body of literature to derive thermodynamic, kinetic, and microstructural quantities to use as inputs to precipitation models. In general, the more practical approach to modeling precipitation kinetics is to tune input parameters of precipitation models to match experimental data. This process can be refined by the use of probabilistic \cite{Honarmandi2020} or optimization techniques \cite{Yu2024}. However, these tuning approaches are built on top of software, such as those used in TC-PRISMA \cite{Chen2014}, MatCalc \cite{Svoboda2004,Kozeschnik2004}, Pandat \cite{Cao2009}, and Kanwin\footnote{Certain equipment, instruments, software, or materials are identified in this paper in order to specify the experimental procedure or concept adequately. Such identification is not intended to imply recommendation or endorsement by the National Institute of Standards and Technology, nor is it intended to imply that the materials or equipment identified are necessarily the best available for the purpose.} \cite{Ury2023}, which adopt model forms that have made many assumptions about precipitation behavior. Thus, tuning of input parameters alone is not sufficient to accurately model precipitation kinetics, as shown by the discrepancies present in Refs. \cite{Honarmandi2020,Yu2024}.

While there are many assumptions related to the thermodynamic, kinetic, and microstructural contributions to precipitation modeling, in this work we will focus on the thermodynamic modeling of coherent precipitates. Specifically, we will focus on two standard assumptions: that elastic effects can be neglected for systems with a small lattice misfit and that size effects can be neglected for low levels of supersaturation. The neglect of elastic effects can be seen in, for example, phase-field based modeling of precipitation in the Fe-Cr system due to the low lattice misfit between the two elements \cite{Li2014, Ke2019, Tissot2023}. In nucleation modeling, the option to directly include elastic effects is usually present in the form of an estimate for the self energy of the precipitate but not for the elastic interaction with the rest of the microstructure. This later contribution can be indirectly included through an energy addition term that accounts for unknown contributions, but this term does not evolve with changes in microstructure while elastic interactions will \cite{Yu2024}. The neglect of size effects is commonly taken in the form of adopting the capillarity approximation, where it is assumed that the interfacial free energy is independent of precipitate size. This assumption has been known to be problematic for smaller sized precipitates for over half a century \cite{Tolman1949}, and yet is commonly  adopted in precipitation modeling. The presence of size effects on the interfacial free energy has been verified through simulations of Ising and Lennard-Jones (L-J) models of droplets and bubbles \cite{Binder2011}, but these models do not require dealing with the complexity of elastic effects such as those present in coherent precipitates.

The above assumptions are especially unwarranted when modeling precipitation in highly supersaturated microstructures, such as those found in additive manufacturing (AM). When a matrix is highly supersaturated, there will be a large precipitate fraction present after nucleation and growth. Before the late stages of growth and coarsening, there will be many precipitates which each have their own elastic field. Thus, the contribution to the thermodynamic description of coherent precipitates from elastic interactions will be large. Additionally, a large supersaturation implies a large thermodynamic driving force for nucleation, which will result in critical nuclei forming at small sizes. The smaller the nuclei the less valid is the capillarity approximation. During AM based processes that rely on rapid solidification, interdendritic regions will have a degree of microsegregation present. A high amount of microsegregation can cause a region of material to be locally supersaturated, as exemplified by an experimental study of Inconel 625 \cite{Lass2017}. The global composition of this alloy is expected to be single phase at the temperatures studied, but when undergoing annealing after the laser powder-bed fusion AM process, many precipitates appeared in the interdendritic regions. The formation of these precipitates was attributed to supersaturation in the interdendritic regions due to the presence of microsegregation. These precipitates are undesirable in Inconel 625 due to their deleterious effects on mechanical properties \cite{Mittra2013}. The ability to design optimal post-processing treatments for AM depends on accurately modeling precipitation kinetics in these microsegregation-induced supersaturated regions.

The present work will assess the contributions of the aforementioned assumptions on the thermodynamic description of coherent precipitates through the use of classical atomistic simulations and atomistically-informed continuum modeling. We will focus on a model Fe-Cr system, as it allows us to highlight the importance of elastic effects even when lattice mismatch is small. Section II discusses the theoretical framework for our analysis, Section III presents the simulation methodology used to inform the framework, Section IV presents the results of our analysis, Section V discusses the significance of those results, and we conclude with Section VI. Our analysis reveals that not only are significant elastic and size effects present, but also reveals the potential importance of often overlooked higher-order effects. 

\section{Theory of Interfacial Equilibrium}

We use an isobaric semi-grand potential description of the thermodynamics of a coherent precipitate as we assume that the nucleation of a precipitate will have a negligible effect on the composition of the matrix. We consider the case where pressure is set to zero as ambient pressure conditions have a negligible effect on solids. The total semi-grand potential $\Omega$ of a system with a matrix with volume $V^m$ and a coherent precipitate with volume $V^p$ and interfacial area $A^p$ in the sharp-interface limit is \cite{Rottman1988}:
\begin{equation} \label{eq:1}
    \Omega = \int_{V^p} \rho (\vec{r})\omega^p(\vec{r}) d\vec{r} + \int_{V^m} \rho (\vec{r})\omega^m(\vec{r}) d\vec{r} + \int_{A^p} \gamma dA
\end{equation}
where $\rho (r)$ is the local atomic density, $\omega (r)$ is the local semi-grand potential per atom, $\gamma$ is an isotropic interfacial free energy, and superscripts $p$ and $m$ denote the precipitate and matrix phases, respectively. To simplify the above equation, we make the following assumptions: the precipitate is spherical with radius $R$, any heterogeneity in $\omega(r)$ within a phase is due to elastic effects, composition-stress coupling can be neglected, and the difference between lattice constants of the precipitate and matrix are small enough such that the atomic density can be approximated as uniform and equal between the two phases. Eq. \ref{eq:1} can then be written as:
\begin{equation} \label{eq:2}
    \Omega = \frac{4}{3}\pi R^3 \rho_0 \omega_0^p + (V-\frac{4}{3}\pi R^3)\rho_0 \omega_0^m + E^{el} + 4\pi R^2 \gamma
\end{equation}
where $\rho_0$ is the uniform atomic density, $\omega_0$ is the semi-grand potential of a phase in its unstressed state, $V$ is the total volume of the system and $E^{el}$ is the total elastic energy of the system.

Taking the derivative on both sides of Eq. \ref{eq:2} with respect to $R$ while assuming that a change in $R$ has a negligible effect on $\rho_0$ results in the following equation:
\begin{equation} \label{eq:3}
    \frac{d\Omega}{dR} = 4\pi R^2 \rho_0 (\omega^p _0 -\omega^m_0) + \frac{\partial E^{el}}{\partial R} + 8\pi R \gamma+4\pi R^2 \frac{\partial\gamma}{\partial R}.
\end{equation}
In the above equation, we have allowed the elastic energy and the interfacial free energy to depend on $R$. Setting the left-hand side of Eq. \ref{eq:3} to zero and rearranging results in:
\begin{equation} \label{eq:4}
    \rho_0 (\omega^m_0 -\omega^p_0)-\frac{\partial E^{el}/\partial R}{4\pi R^2} = \frac{2\gamma}{R}+\frac{\partial\gamma}{\partial R}.
\end{equation}
Eq. \ref{eq:4} is the condition that must be satisfied for the maximum of Eq. \ref{eq:2}. This is the condition of equilibrium at the interface of a spherical precipitate that will neither grow nor shrink, which is also known as the critical nucleus condition. As will be discussed in detail in the next section, the first term on the left-hand side of Eq. \ref{eq:4} will be determined from atomistic simulations, the second term through atomistically-informed continuum modeling, and $\gamma(R)$ can then be extracted by parameterizing an appropriate size-dependent interfacial free energy function. While we have made a few assumptions in its derivation, Eq. \ref{eq:4} allows us to study the relative importance of elastic effects and the size-dependence of the interfacial free energy.

The value of the interfacial free energy in the presence of finite curvature is dependent on the choice of a dividing surface which determines the position of the interface as well as the radius of the precipitate. As long as the total semi-grand potential of the system remains constant, the dividing surface can, in principle, be placed at any arbitrary location, although some choices are certainly more convenient for analysis. We choose to work with the equimolar dividing surface, which represents the location at which the interface is defined to have zero excess solute. This allows us to take advantage of the lever rule when determining phase fractions and phase compositions \cite{Trster2012} as discussed below. When choosing a different dividing surface, the determination of phase fractions and compositions becomes more complicated. Additionally, different dividing surfaces can result in equations slightly different than Eqs. \ref{eq:3} and \ref{eq:4} above due to changes in the last term on the right-hand sides of those equations.

\section{Methods}

We model a body-centered cubic (BCC) Fe-Cr system using the same concentration-dependent embedded atom method (CD-EAM) interatomic potential that was developed in Ref. \cite{Abu-Odeh2025}. The results in this work are not meant to be quantitatively representative of the true Fe-Cr system, but the potential that we are using allows us to explore precipitate thermodynamics in the limit of a small lattice misfit (below 1 \%). Thus, the results are expected to provide important qualitative insights to the thermodynamic modeling of coherent precipitates. All atomistic simulations are carried out using the Large-scale Atomic/Molecular Massively Parallel Simulator (LAMMPS) software \cite{Plimpton1995}. Many of the methods used in this study mimic that of our previous work in Ref. \cite{Abu-Odeh2025}, so we refer the interested reader to that study for further details. The remainder of this section will briefly review the methods used in Ref. \cite{Abu-Odeh2025}, and elaborate on the methods unique to this work.

Hybrid Monte Carlo/molecular dynamics (MC/MD) simulations using fully periodic boundary conditions are carried out to determine the semi-grand potential of Fe-rich phase $\alpha$ and the Cr-rich phase $\alpha '$. The semi-grand potential of a phase is defined as:
\begin{equation} \label{eq:5}
    \omega (c) = g(c) - c\Delta\mu(c)
\end{equation}
where $g$ is the Gibbs free energy per atom, $c$ is the atomic concentration of Cr, and $\Delta \mu$ is the diffusion potential, which is equal to the difference in chemical potentials of Cr and Fe. Thermodynamic quantities are obtained using a mix of the isobaric semi-grand canonical (SGC) and the isobaric variance-constrained semi-grand canonical (VC-SGC) ensembles \cite{Sadigh2012a,Sadigh2012b}. Modifications must be made to the treatment of these ensembles in LAMMPS to obtain quantitatively correct thermodynamic properties \cite{Abu-Odeh2025}. $\Delta \mu(c)$ is directly obtained through hybrid MC/MD. The parameterization of a function representing $g(c)$ requires both the integration of $\Delta \mu(c)$ and the Gibbs free energy of the end members (pure Fe and Cr), the latter of which is obtained using a nonequilibrium thermodynamic integration approach \cite{Freitas2016}. Since we are fixing the total pressure of the system to be zero, the Gibbs free energy can be replaced with the Helmholtz free energy without any change in the results of this work.

MC/MD in the VC-SGC ensemble is used to sample states with a precipitate embedded in a matrix. The VC-SGC ensemble allows for determination of $\Delta \mu (c)$ within the two-phase region of the phase diagram while suppressing global compositional fluctuations about a mean value. As we are adopting an equimolar dividing surface and are neglecting stress-composition coupling, we are able to estimate both the phase fraction and the phase compositions using the lever rule \cite{Trster2012}. For a given precipitate-matrix state at a composition $c$ with a diffusion potential of $\Delta \mu (c)$, there will be a composition $c_\alpha$ of the $\alpha$ phase and a composition $c_{\alpha '}$ of the $\alpha '$ phase with the same value of $\Delta \mu$. According to the lever rule, the phase fraction of $\alpha$, $f_\alpha$, is given by:
\begin{equation} \label{eq:6}
    f_\alpha = \frac{c_{\alpha '}-c}{c_\alpha - c_{\alpha '}}
\end{equation}
and the phase fraction of $\alpha '$, $f_{\alpha '}$, is given by:
\begin{equation} \label{eq:7}
    f_{\alpha '} = \frac{c-c_{\alpha}}{c_\alpha - c_{\alpha '}}.
\end{equation}
Once the precipitate phase fraction is determined, a precipitate radius can be estimated from:
\begin{equation} \label{eq:8}
    R = (f_p V\frac{3}{4\pi})^{1/3}
\end{equation}
where $V$ is the average volume of the supercell containing the precipitate and the matrix during the MC/MD simulation. Eq. \ref{eq:8} assumes a uniform atomic density within the supercell. The atomic density $\rho_0$ is obtained by dividing the total number of atoms in the supercell with the precipitate-matrix configuration by $V$. To sample a wide range of precipitate sizes, we use simulation box sizes ranging from 20 $\times$ 20 $\times$ 20 BCC unit cells to 60 $\times$ 60 $\times$ 60 BCC unit cells (16,000 to 432,000 atoms). All MC/MD simulations had a pre-equilibration period of 2000 MC sweeps (where one MC sweep is completed when the number of MC attempts equals the total number of atoms in the simulation cell) with statistics taken from the subsequent 4000 MC sweeps. Ten MD steps were taken after each MC sweep using a time step of 2.5 fs.

To determine the elastic contribution, we make the following estimate:
\begin{equation} \label{eq:9}
    \frac{\partial E^{el}}{\partial R} \approx \frac{E^{el}(R+q)-E^{el}(R-q)}{2q}
\end{equation}
where we set $q$ to 0.1 nm. $E^{el}$ is estimated from a continuum anisotropic linear elasticity model as discussed in Ref. \cite{Abu-Odeh2025}. We found that the continuum elasticity model is able to closely replicate the topology and magnitude of the finite-temperature stress fields associated with a coherent precipitate when compared to atomsitic simulation data. The continuum model requires atomistic input such as concentration-dependent finite-temperature elastic constants, concentration-dependent lattice constants, and an estimate for the concentration field. We use the previously obtained values of $c_\alpha$, $c_{\alpha '}$, and $R$ to generate a continuum concentration field. This is done through the use of the following function:
\begin{equation} \label{eq:10}
    c(r) = \frac{c^m + c^p}{2}+(c^m - \frac{c^m + c^p}{2})\mathrm{tanh}(\frac{r-R}{l})
\end{equation}
where $r$ is the radial distance away from the center of the precipitate and $l$ is a scaling parameter related to the interface width. While we previously used measured interfacial widths to parameterize $l$ \cite{Abu-Odeh2025}, in this work we fix $l$ to the small value of 0.05 nm. This is due to our desire to maintain a sharp interface picture of coherent precipitates, and use of finite interface widths can result in excess elastic energy in comparison with analytical sharp interface solutions \cite{Simon2020}. As we are using a spectral solver for our continuum model, we still opt to have a finite value of $l$ to avoid spurious effects of sharp interfaces with spectral solvers. We use a grid spacing of half the effective lattice constant of the atomistic simulation cell for the spectral solver, which naturally supports fully periodic boundary conditions.

Using the information obtained above, we can now estimate $\gamma(R)$. To do so, we assume that $\gamma(R)$ behaves according to the following function \cite{Binder2011}:
\begin{equation} \label{eq:11}
    \gamma(R) = \frac{\gamma(\infty)}{1+2\delta/R+2(\lambda/R)^2}
\end{equation}
where $\gamma(\infty)$ is the interfacial free energy in the limit of no curvature (i.e., a flat interface), $\delta$ is usually referred to as the Tolman length \cite{Tolman1949}, and $\lambda$ accounts for the $R$ dependence of $\delta$. We treat all three constants as fitting parameters which best match the right-hand side of Eq. \ref{eq:4} to its left-hand side.

We provide here an important comment on the use of the VC-SGC ensemble to extract interfacial free energies from thermodynamic integration as discussed in Ref. \cite{Sadigh2012a}, which we will refer to as SE for Sadigh and Erhart, as opposed to using Eq. \ref{eq:4} as is done in this work. In SE, the total free energy (minus a constant that is not important for this discussion) is obtained through the integration of a $\Delta\mu (c)$ versus $c$ curve through the single phase regions, the spherical precipitate regions, the cylindrical precipitate regions, and the slab regions. SE does this for a Fe-Cr system with system sizes below 500 atoms using fully periodic boundary conditions. From our preliminary attempts at using this method for much larger system sizes, we found appreciable hysteresis effects at transition points between a single phase to a spherical precipitate, a spherical precipitate to a cylindrical precipitate, and a cylindrical precipitate to a slab configuration. These hysteresis effects make direct integration of $\Delta\mu (c)$ unreliable as the resulting free energy values will be dependent on the path of integration. Previous work using a successive umbrella sampling technique (which is similar in spirit to MC in the VC-SGC ensemble) to probe the transition of a homogeneous L-J gas into a liquid droplet found that near the evaporation/condensation transition exist probability distributions of internal energy and chemical potential at multiple densities that result in two distinct peaks \cite{MacDowell2004,Schrader2009}. This is indicates the existence of a barrier between the two states where it could be difficult to determine the exact transition point when using MC for a finite amount of steps. This would be analogous to a barrier between the homogeneous solid phase and the state with a spherical coherent precipitate. We believe such a barrier would also exist between the spherical precipitate state and the cylindrical state, and between the cylindrical state and the slab state. This barrier is likely the cause of the hysteresis effects we previously observed. We believe that SE did not report any hysteresis effects because their system sizes were so small that spatially local compositional fluctuations could easily find their periodic images to transition into a different state. The reduction of transition barriers when using smaller simulation sizes with periodic boundary conditions has also been observed when using umbrella sampling to simulate lipid membrane pore closing \cite{Awasthi2016}. Thus, unless one is using extremely small system sizes, we advise against the direct integration method of VC-SGC results as presented in SE until a reliable method to deal with hysteresis effects is found.

\section{Results}

\begin{figure}[h!]
    \centering
    \includegraphics[width=1\textwidth]{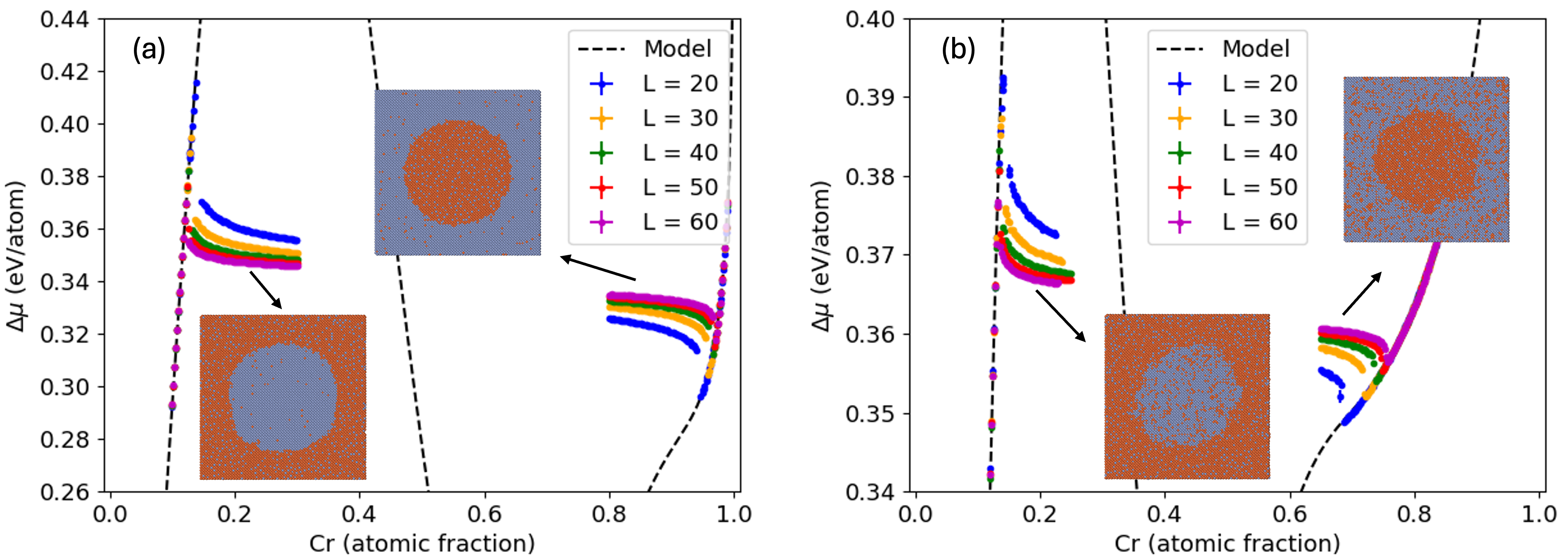}
    \caption{Results from hybrid MC/MD for (a) 600 K and (b) 1000 K. The different color points represent data from different simulation box sizes, with $L$ representing the number of unit cells across a box's length. The error bars (which are smaller than the data points) represent +/- two times the standard error. The black dashed line represents a thermodynamic model describing the homogeneous phase data. The insets show representative atomistic snapshots of precipitate microstructures in the largest box size.}
    \label{fig:mc_data}
\end{figure}

Figure \ref{fig:mc_data} presents the results obtained directly from hybrid MC/MD simulations. The simulations provide data representing the relationship between $\Delta\mu$ and $c$ for the homogeneous single phases as well as for the two-phase mixture of a spherical precipitate in a larger matrix. The black dashed lines in Figure \ref{fig:mc_data} represent a thermodynamic model to describe the data of the homogeneous phases (more details of the model are given in Ref. \cite{Abu-Odeh2025}). The thermodynamic data relating to spherical precipitates are present in the data in the middle of the figures in between the side branches of the homogeneous phase data. It can be seen that there are regions in the figure where a given $\Delta\mu$ value can be found at three different compositions of Cr. The lever rule (Eqs. \ref{eq:6} and \ref{eq:7}) is applied in these regions to determine the phase fraction and composition of the phases in the two-phase microstructure. With the phase compositions, the semi-grand potentials are available through Eq. \ref{eq:5}, where the value of $g(c)$ for each phase is determined through the integration of the thermodynamic model. This also enables an estimate of the precipitate radius through Eq. \ref{eq:8}.

We additionally use the data from the hybrid MC/MD simulations to estimate the level of supersaturation of the matrix phase explored in this work. The composition of the supersaturated matrix is compared to the coexistence composition determined from the common-tangent construction from Ref. \cite{Abu-Odeh2025}. While this construction assumes the two phases to be incoherent, we use the results for the sake of comparison. We use two measures of supersaturation: the absolute difference between the supersaturated composition and the one at coexistence ($|c_{ss}-c_{co}|$), and the absolute difference between one and the ratio of the two compositions ($|1-c_{ss}/c_{co}|$). The supersaturation levels are reported in Table \ref{tab:supersat}. Across temperatures and between the different matrices, the supersaturation in this work ranges from 0.1 \% to 9 \%, depending on the method used. Both methods show that there are supersaturation levels below 1 \% being explored. As we will see later in the section, even with these relatively small amounts of supersaturation, there is a size-dependence for the interfacial free energy.

\begin{table}[h!]
    \centering
	\begin{tabular}{l | l | l | l | l} 
        Matrix & $c_{co}$ & $c_{ss}$ & $|c_{ss}-c_{co}|$ & $|1-c_{ss}/c_{co}|$ \\ \hline 
        $\alpha$ 600 K & 0.113 & 0.115-0.123 & 0.002-0.01 & 0.018-0.088 \\ \hline
	$\alpha'$ 600 K & 0.982 & 0.967-0.98 & 0.002-0.015 & 0.002-0.015 \\ \hline
        $\alpha$ 1000 K & 0.127 & 0.128-0.133 & 0.001-0.006 & 0.008-0.047 \\ \hline
        $\alpha'$ 1000 K & 0.801 & 0.729-0.787 & 0.014-0.072 & 0.017-0.09 \\

    \end{tabular}
	\caption{\label{tab:supersat} Coexistence and supersaturated compositions for the matrix phases along with two different measures of relative supersaturation. The last three columns represent the minimum and maximum values over the range of supersaturated compositions. Definitions and discussion of these variables are provided in the main text.}
\end{table}

\begin{figure}[h!]
    \centering
    \includegraphics[width=1\textwidth]{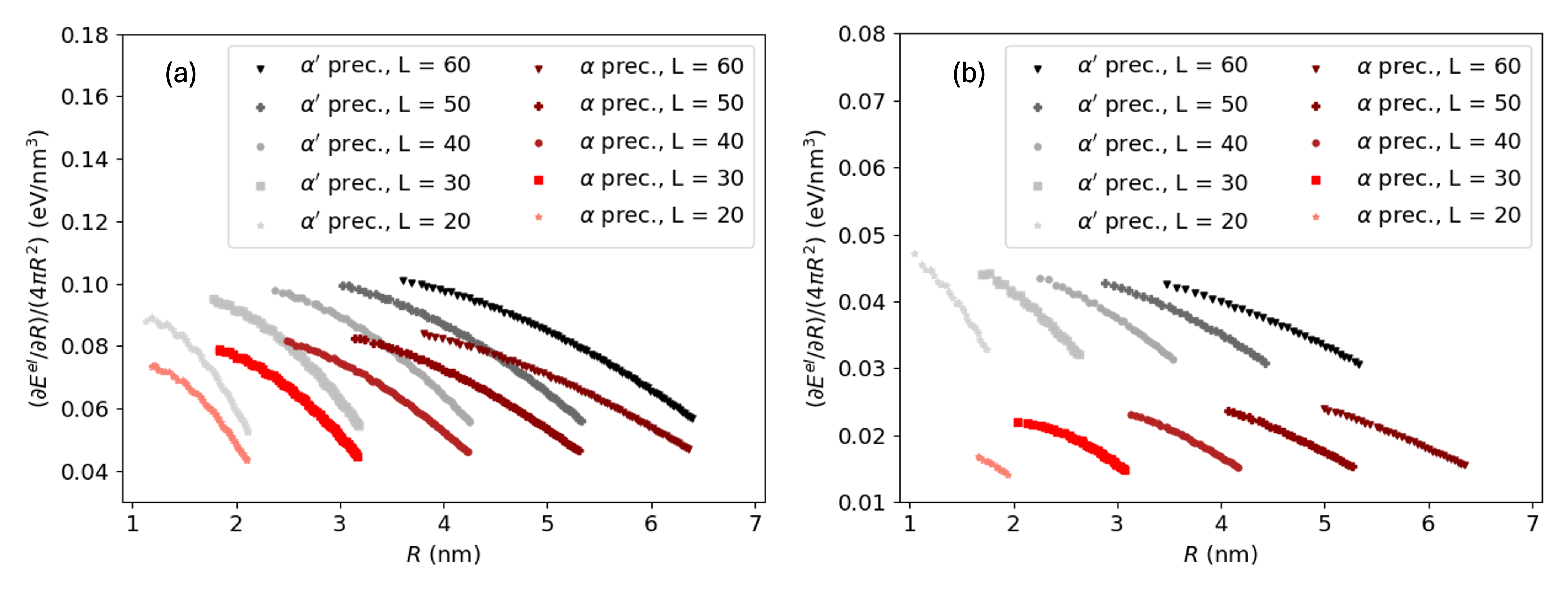}
    \caption{Change in the elastic energy of the system with a change in precipitate radius divided by the interface area at (a) 600 K and (b) 1000 K. The gray/black data points are values for systems with $\alpha'$ precipitates and the red data points are values for systems with $\alpha$ precipitates. The different shades/symbols correspond to values for different system box sizes.}
    \label{fig:elastic_data}
\end{figure}

The previously determined phase compositions and precipitate radii are used in the atomistically-informed continuum elasticity model \cite{Abu-Odeh2025}. The elastic term in Eq. \ref{eq:4} is calculated using Eq. \ref{eq:9} and its value is plotted in Figure \ref{fig:elastic_data} for different precipitates, radii, simulation box sizes, and temperatures. Figure \ref{fig:elastic_data} shows that for a given radius and a given box size, the elastic contribution for the $\alpha'$ precipitate is larger than the $\alpha$ precipitate. This is expected due to the fact that the $\alpha'$ phase has larger elastic constant values than the $\alpha$ phase. Additionally, it can be seen that a given precipitate radius will have a different elastic contribution at a different box size. This has to do with the fact that fully periodic boundary conditions are in place. The precipitate in the simulation box is interacting with an infinite array of its own images, with the interaction at a fixed radius being stronger with a smaller box size as the images are closer to the original precipitate.

\begin{figure}[h!]
    \centering
    \includegraphics[width=1\textwidth]{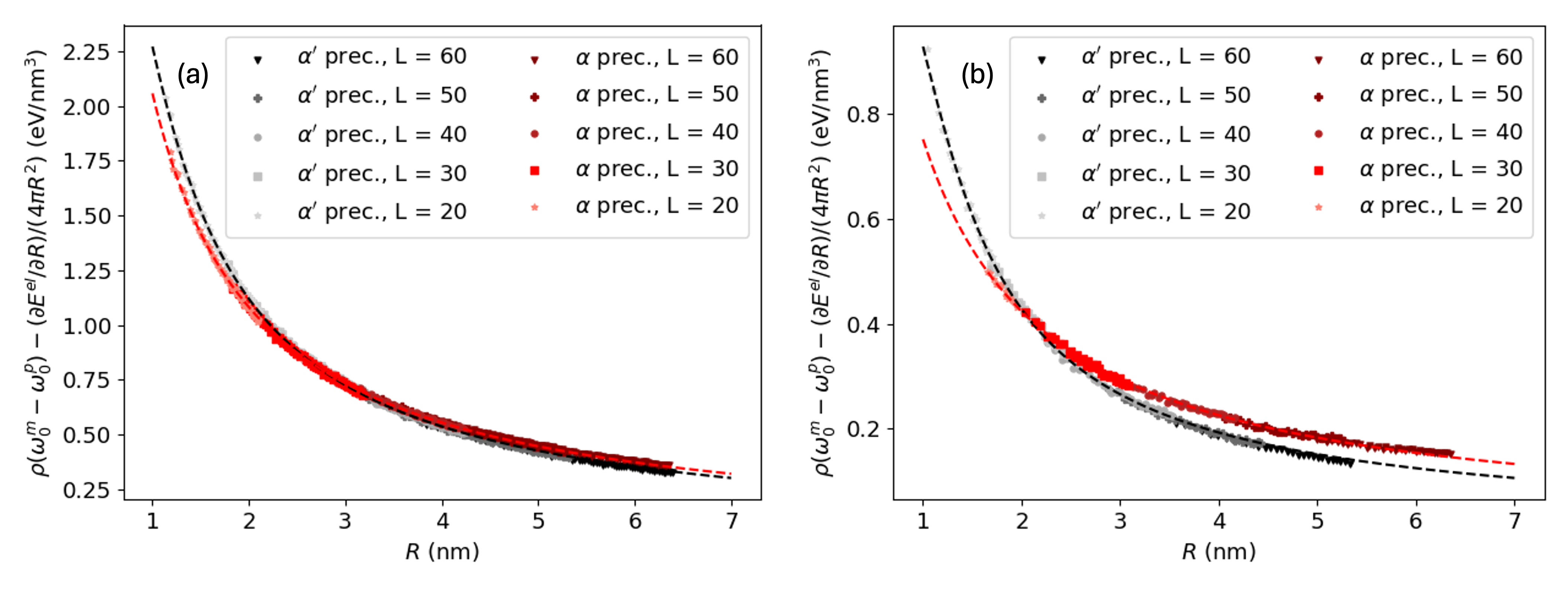}
    \caption{Volumetric driving force as a function of radius at (a) 600 K and (b) 1000 K. The gray/black data points are values for systems with $\alpha'$ precipitates and the red data points are values for systems with $\alpha$ precipitates. The different shades/symbols correspond to values for different system box sizes. The dashed lines represent the right-hand side of Eq. \ref{eq:4} after obtaining the parameters for Eq. \ref{eq:11}.}
    \label{fig:driving_force}
\end{figure}

Figure \ref{fig:driving_force} plots the values of the left hand side of Eq. \ref{eq:4} for different precipitates, precipitate radii, and temperatures. We will refer to this total value as the volumetric driving force. Without the elastic contribution, there would be a clear separation of values of the volumetric driving force for the same radius at different box sizes for the same type of precipitate. This is due to periodic boundary effects, which are well captured by the continuum elasticity model. As the radius decreases, for both temperatures, the volumetric driving force for the $\alpha'$ precipitate becomes larger than that for the $\alpha$ precipitate. Conversely, as the radius increases the volumetric driving force for the $\alpha'$ precipitate becomes smaller than that for the $\alpha$ precipitate. This will affect the trend of the interfacial free energy as a function of radius for each precipitate.

\begin{figure}[h!]
    \centering
    \includegraphics[width=1\textwidth]{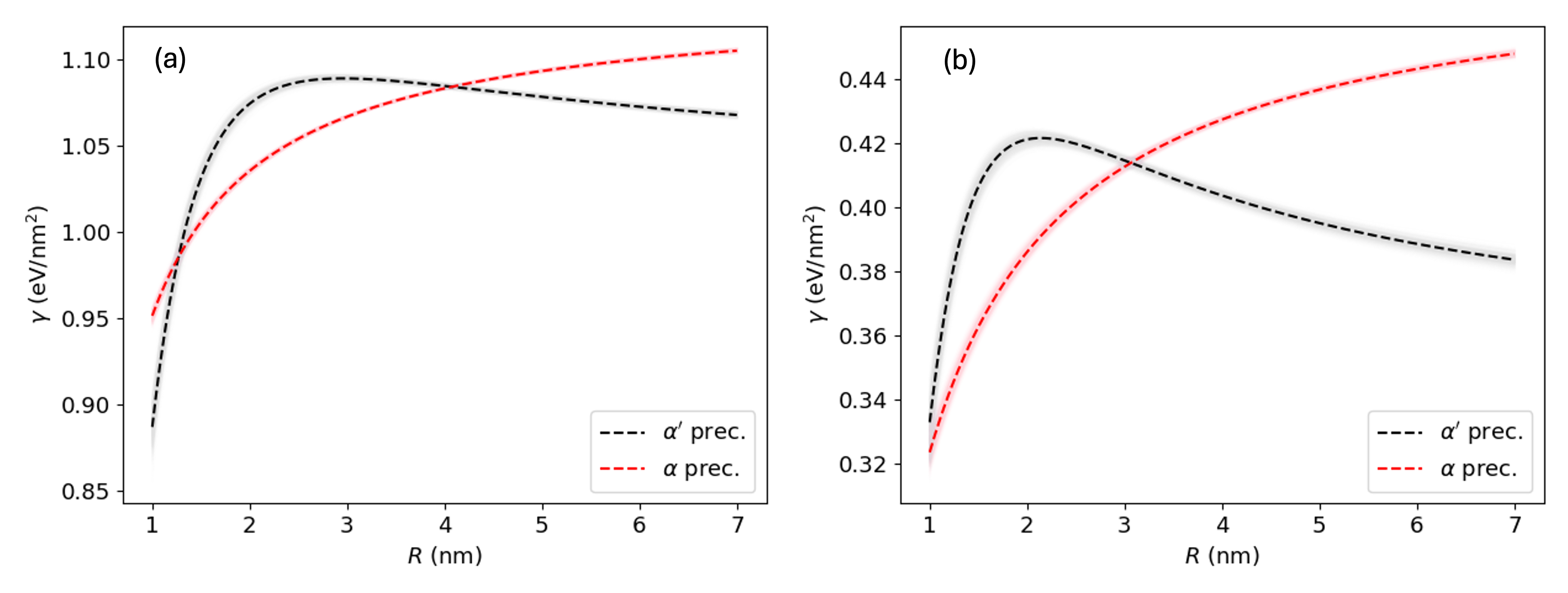}
    \caption{Interfacial free energies for precipitates at different sizes at (a) 600 K and (b) 1000 K. Shaded regions represent an estimated uncertainty based on randomly drawing parameter sets from a multivariate Gaussian distribution with a covariance matrix obtained from the best fit of Eq. \ref{eq:11} to the data in Fig. \ref{fig:driving_force}.}
    \label{fig:interfacial_free}
\end{figure}

Eq. \ref{eq:4} combined with Eq. \ref{eq:11} allows for the extraction of values of $\gamma(\infty)$, $\delta$, and $\lambda$ from the volumetric driving force data through the use of a non-linear least-squares regression. These values are given in Table \ref{tab:int_para}. The resulting values of $\gamma(R)$ are plotted in Figure \ref{fig:interfacial_free}. The $\lambda$ parameter is not reported for the $\alpha$ precipitate as the value obtained from the regression did not have much of an effect on the model when compared to using zero. Additionally, we note that the regression found another parameter set for $\delta$ and $\lambda$ for the $\alpha'$ precipitate when the initial guess for the parameters is changed, but it led to nonphysical behavior, such as negative interfacial free energies, when extrapolating to even smaller radii. It is immediately clear when looking at Figure \ref{fig:interfacial_free} that the capillarity approximation does not hold within the range of precipitate sizes investigated. To highlight the importance of considering elastic effects, we found that if we were to completely neglect elasticity in the volumetric driving force that the obtained values of $\gamma(\infty)$ would increase by approximately 14 \% - 23 \%. As discussed earlier, such a change in interfacial free energy values can yield dramatic changes in precipitation kinetics \cite{Tavenner2024}. Both $\delta$ and $\lambda$ show an increase in magnitude with an increase in temperature. Additionally, a change in sign of $\delta$ is dependent on whether the interface is for an $\alpha$ or $\alpha'$ precipitate. The temperature dependence of these parameters as well as the change in sign of $\delta$ with a change in precipitate type is in agreement with trends of L-J droplets and bubbles \cite{Block2010}. However, an unexpected result from the present work is that for a given temperature the obtained value of $\gamma(\infty)$ is different for each precipitate. Looking at results from L-J droplets and bubbles, $\gamma(R)$ is expected to converge to the same value at infinitely large $R$ \cite{Block2010}. The main difference between this work and the work on L-J systems is that the current work is focused on coherent solid phases as opposed to fluid phases. We believe that the origin of the different values of $\gamma(\infty)$ for different precipitates is associated with higher order effects related to elasticity.

\begin{table}[h!]
    \centering
	\begin{tabular}{l | l | l | l } 
        Precipitate & $\gamma(\infty)$ (eV/nm$^2$) & $\delta$ (nm) & $\lambda$ (nm) \\ \hline 
        $\alpha'$ 600 K & 1.028(2) & -0.166(4) & 0.495(9) \\ \hline
	$\alpha$ 600 K & 1.136(1) & 0.097(2) & $-$ \\ \hline
        $\alpha'$ 1000 K & 0.350(2) & -0.36(1) & 0.62(1) \\ \hline
        $\alpha$ 1000 K & 0.479(2) & 0.240(8) & $-$ \\

    \end{tabular}
	\caption{\label{tab:int_para} Parameters for Eq. \ref{eq:11} obtained through non-linear least-squares regression. The $\lambda$ parameter is not reported for the $\alpha$ precipitate as it was very small in magnitude and did not have an appreciable effect. Error estimates of the parameters are given in the parentheses.}
\end{table}

\begin{figure}[h!]
    \centering
    \includegraphics[width=1\textwidth]{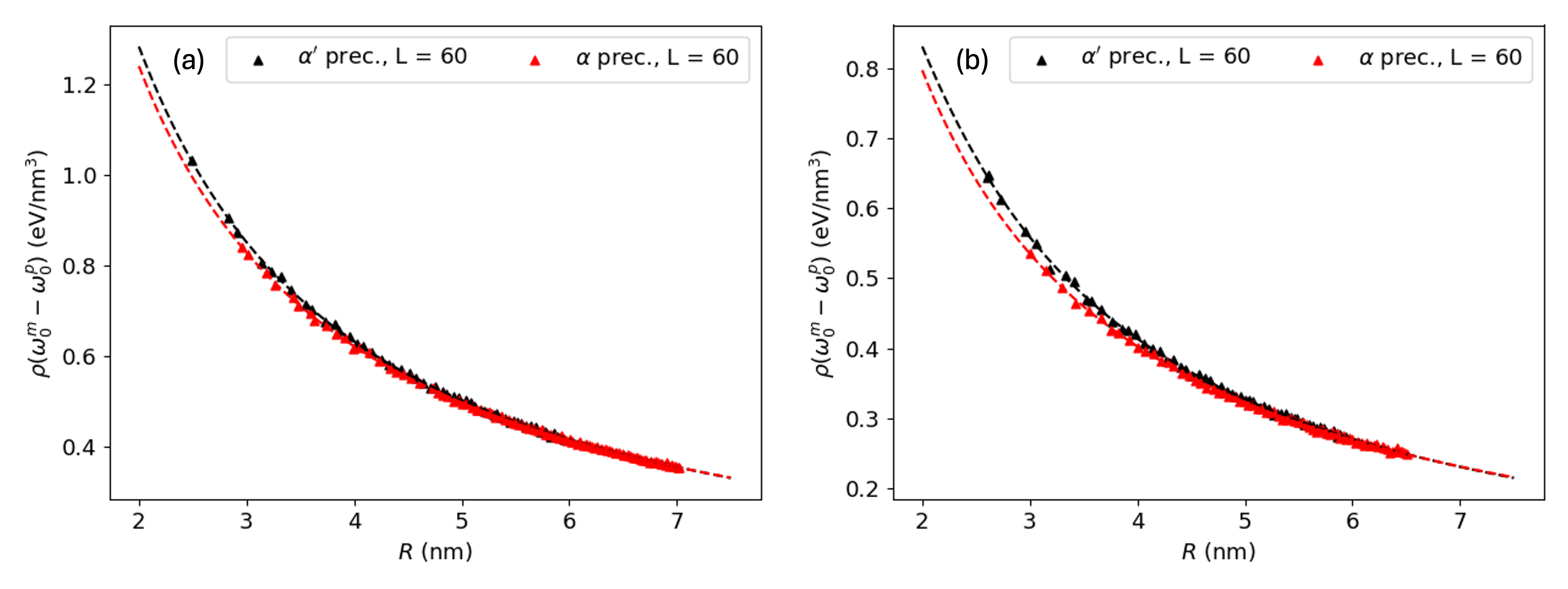}
    \caption{Volumetric driving force as a function of radius at (a) 600 K and (b) 1000 K for the case were only MC is used at a fixed lattice spacing and box size. The dashed lines represent the right-hand side of Eq. \ref{eq:4} after obtaining the parameters for Eq. \ref{eq:11}.}
    \label{fig:no_md}
\end{figure}

To check that these effects are due to the presence of elasticity, we carry out further simulations only using MC for a fixed lattice spacing and a fixed box size. This removes the possibility for displacement gradients and volume relaxation to occur. We only focus on the larger precipitate sizes in this case, so we limit MC simulations to boxes which are 60 $\times$ 60 $\times$ 60 BCC unit cells. We use the same analysis as was done for the hybrid MC/MD case except, elasticity is ignored. The results are given in Figure \ref{fig:no_md}. Similar to the case where MD is used, as the radius decreases, at both temperatures, the volumetric driving force for the $\alpha'$ precipitate becomes larger than that for the $\alpha$ precipitate. Unlike the case where MD is used, as the radius increases the volumetric driving force for the two precipitates converge with each other. Additionally, the obtained values of $\gamma(\infty)$ are 1.221(7) eV/nm$^2$ and 1.249(2) eV/nm$^2$ for $\alpha'$ and $\alpha$ precipitates at 600 K, and are 0.781(7) eV/nm$^2$ and 0.814(2) eV/nm$^2$ for $\alpha'$ and $\alpha$ precipitates at 1000 K. The values of $\gamma(\infty)$ for different precipitates are much closer to each other than when elasticity is present as seen in Table \ref{tab:int_para}. These results support that the origin of why the volumetric driving forces in Figure \ref{fig:driving_force} do not converge to the same value, as well as why the values of $\gamma(R)$ do not converge to the same value in Figure \ref{fig:interfacial_free}, is related to elastic effects.

\section{Discussion}

The results of this work reveals two critical insights. The first is that size effects on interfacial free energies are significant, and should not be treated as negligible,  even for relatively small supersaturations. This is evidenced by Figure \ref{fig:interfacial_free}. The precipitates at larger values of $R$ represent precipitates in a matrix with a supersaturation value of around 1 \% or less. If finite size effects were negligible, then the curves in Figure \ref{fig:interfacial_free} would be horizontal lines. At larger supersaturations, where the critical nucleus size is smaller, the size effects are more pronounced. The extracted interfacial free energies can be reduced by approximately 10 \% to 30 \% of the value of $\gamma(\infty)$ depending on the precipitate size, type, and temperature. As shown in Ref. \cite{Tavenner2024}, such a change can dramatically affect predicted precipitation kinetics. Thus, to improve the physical accuracy of precipitation models such as those in Refs. \cite{Chen2014,Svoboda2004,Kozeschnik2004,Cao2009,Ury2023}, it is necessary to account for the effect of curvature on the interfacial free energy using a model such as Eq. \ref{eq:11}. The online documentation for MatCalc and Pandat discusses the inclusion of a size dependent function that alters the interfacial free energy based on a bond-breaking model \cite{Sonderegger2009}. However, the interfacial free energy in this function monotonically decreases with a decrease in size, and the size-dependent terms are independent of temperature. Both of these features are at odds with the results of the present work as well as the results of previous studies on L-J fluids \cite{Block2010}. The online documentation for TC-PRISMA discusses the ability for the user to define temperature/size-dependent interfacial free energy functions, but does not provide any recommendations for the form of the functions. We suggest that any precipitation kinetics software simulators use Eq. \ref{eq:11} to capture size-effects on the interfacial free energy as this will significantly influence nucleation and early growth kinetics when precipitate sizes are small.

The second insight uncovered by this effort is that elastic effects are not negligible, even for phases with a relatively small lattice mismatch. Even though the lattice mismatch between the two phases in this work is less than 1 \%, the elastic energy contribution was found to account for approximately 2.5 \% to 25 \% of the total volumetric driving force depending on the precipitate type and size. Neglecting this contribution was shown to have a strong impact on the determination of the interfacial free energy. Based on Figure \ref{fig:elastic_data}, the elastic contribution is expected to be dampened in this work due to the interaction with periodic images. Thus, the elastic contribution for systems where the precipitate fraction is more dilute is expected to be even larger than that found in this work. We suggest that elastic effects should not be neglected even when dealing with coherent precipitates with low lattice misfit. This would not only affect traditional precipitate kinetics models, but also phase-field based modeling \cite{Li2014,Ke2019,Tissot2023}. While precipitate models such as TC-PRISMA, MatCalc, Pandat, and Kanwin allow for the inclusion of a volumetric elastic contribution, the effect of elastic interactions between precipitates are not accounted for. Based on Figure \ref{fig:elastic_data} elastic interactions are important even for precipitates with a small lattice mismatch, but the use of the elastic modeling technique in this work is expected to be too computationally expensive for practical precipitate kinetics modeling. When assuming homogeneous elastic constants, an elastic interaction energy can be efficiently calculated provided a spatial distribution of precipitates is given \cite{Shen2006}. This would require the development of an inexpensive way to extend standard precipitation models to track the evolution of the spatial correlation of microstructural features.

To our knowledge, our work extends the previous literature on extracting thermodynamic quantities of coherent precipitates from atomistic simulations by including both size- and elastic-effects. Previous atomistic simulations which have claimed to extract interfacial free energies from coherent precipitates in the Fe-Cr \cite{Sadigh2012a} and the Ni-Al \cite{Tavenner2024} systems have done so for small precipitate sizes compared to those in this work. They did not consider how size effects would change the interfacial free energy nor did they consider the contribution of elastic effects. The present work builds on their efforts to show the importance of those effects.

Even though we consider both size-effects and elastic contributions in our analysis, that the two curves at each temperature in Figure \ref{fig:interfacial_free} do not converge to the same value at large $R$ is surprising and suggests that higher-order effects are at play. We have made a number of assumptions in our analysis, and understanding the consequences of these assumptions is necessary to determine what higher-order effect is important. We itemize these assumptions below, the majority of which require further study to determine their relative importance.

\begin{figure}[h!]
    \centering
    \includegraphics[width=0.5\textwidth]{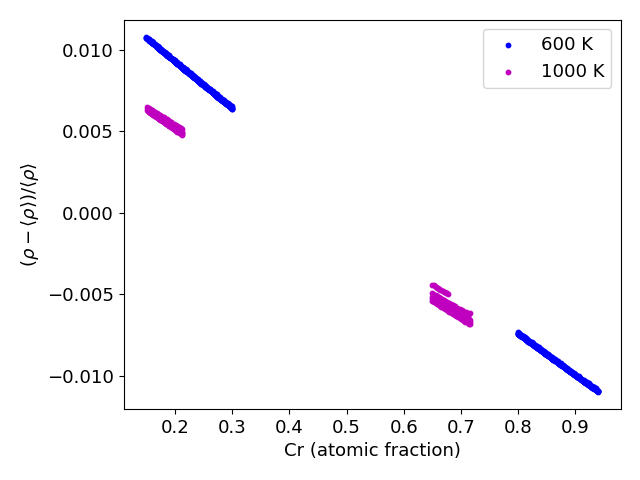}
    \caption{Relative atomic density as a function of composition compared to the mean density at a given temperature.}
    \label{fig:density}
\end{figure}

First, we assumed that the density can be treated as homogeneous and that the change in the density as a function of radius is negligible. Figure \ref{fig:density} shows the measured density at an average global composition relative to the average density $\langle \rho \rangle$ for each temperature. Results from all box sizes are present in the figure. Different global compositions of the two-phases and different box sizes result in a change in the precipitate radii. It is clear in Figure \ref{fig:density} that the variation in density at each temperature is extremely small, such that it is reasonable to assume a homogeneous density and to neglect the change in density with respect to a change in precipitate radius. For systems that have a larger lattice mismatch, it may not be reasonable to adopt the same assumptions.

Second, we have assumed that stress-composition coupling is negligible. The consequences of stress-composition coupling are discussed in the literature (see for example Refs. \cite{Voorhees2004,Mishin2015}). The local compositions of each phase may be different than that obtained from the lever rule due to a local stress contribution to the diffusion potential. This effect may cause a change in the estimated radius of the precipitate as well as a change in the volumetric driving force. Additionally, a change in the estimated radius may mean that our choice of an equimolar dividing surface may not be a valid one, and that there may be a thermodynamic contribution of excess solute on the interface. Assessing the effects of stress-composition coupling would require further study using methods that can naturally account for these effects, such as phase-field modeling.

Third, we have assumed that the effect of interfacial stresses is negligible. Theories of equilibrium at coherent interfaces have accounted for the effect of interfacial stresses \cite{Cahn1982,Rottman1988}. An additional energetic contribution is present that is proportional to the interfacial stress multiplied by the strain at the interface. To the authors knowledge, there is not a reliable method to extract the interfacial stresses of a spherical precipitate from atomistic simulations. However, there is an approach in the literature, albeit an involved one, where the interfacial stresses can be estimated from atomistic simulations of a flat interface \cite{Frolov2012a,Frolov2012b}. Future work could use this approach to obtain an estimate of the interfacial stresses as input to the theories of Refs. \cite{Cahn1982,Rottman1988}.

Finally, we have assumed that the precipitate is perfectly spherical. If the interfacial energy is strongly anisotropic, this would cause the precipitate to adopt a different shape. Additionally, even if the interfacial free energy is isotropic, elastic effects can cause the precipitate to adopt a wide variety of shapes \cite{Thompson1994}. The deviation from a spherical shape in three-dimensions or a circular shape in two-dimensions is due to the competition between the volumetric elastic energy and the areal interfacial energy. The softer the elastic constants of the precipitate relative to that of the matrix, the more dramatic the change in shape of the precipitate \cite{Leo1989,Schmidt1997}. It is not obvious based on the atomistic snapshots of our simulations that there is a significantly non-spherical shape present. However, it is clear that the interfaces in our simulations are rough. This means that the actual interfacial area is larger than that of a smooth sphere. If the interfacial roughness is larger for the $\alpha$ precipitate than the $\alpha'$ precipitate, then this may explain the behavior in Figure \ref{fig:interfacial_free} at large $R$. Interestingly, the $\alpha$ phase has smaller elastic constants than the $\alpha'$ phase \cite{Abu-Odeh2025}. This reduces the $\alpha$ precipitate's stability with respect to shape fluctuations from a perfect sphere \cite{Leo1989,Schmidt1997}. To assess the contribution of the possible enhanced interface roughness due to elastic effects would require further detailed study characterizing shape and interface fluctuations about the equilibrium mean shape.

\section{Conclusion}

We have carried out a combination of atomistic simulations and atomistically-informed continuum modeling for a model coherent Fe-Cr system. By analyzing the results using a theory of interfacial equilibrium, we find that even when lattice misfits are below 1 \%, there are substantial elastic effects on the thermodynamic modeling of coherent precipitates. Additionally, we find that even when supersaturation values are below 1 \%, there are substantial size-effects on the interfacial free energy. We have provided suggestions towards future work to improve traditional precipitation modeling tools to account for these effects. Additionally, our analysis reveals the importance of possible higher order effects in the thermodynamic modeling of coherent precipitates. We have identified possible future directions for study to fully elucidate the different contributions, which include assessing the importance of stress-composition coupling effects, the magnitude of the interfacial stress contributions, and the significance of shape/interfacial fluctuations of the precipitate.

\section*{Acknowledgments}
The authors acknowledge insightful discussions with Yuri Mishin, Carelyn Campbell, and Peter Voorhees. A.A. acknowledges support from a NRC Postdoctoral Fellowship while at the National Institute of Standards and Technology, USA.



\end{document}